\documentstyle[12pt,twoside]{article}
\input epsf

\title{\Large\bf QCD SUM RULES AND THE DETERMINATION OF LEADING TWIST NON--SINGLET OPERATOR MATRIX ELEMENTS}
\author 
{
\it \bf  N. Chamoun$$\\
\small  Department of Theoretical Physics, University of Oxford,\\
\small 1-4 Keble Road, Oxford, OX1 3NP, United Kingdom. \\
}		
\date{}

\begin{document}
\maketitle
\begin{abstract}
We use QCD sum rules to determine directly the leading--twist
non--singlet operator matrix elements based on calculations of
three--point correlator functions in configuration space. We find a
different result from that obtained by integrating the structure
functions' expressions obtained by Belyaev and Ioffe based on
calculations of four--point correlators in momentum space. The origin 
of this discrepancy remains unclear.  
\end{abstract}

\section{Introduction}
\begin{figure}
  \epsfxsize=5.cm
  \epsfysize=4.cm
  \centerline{\epsfbox{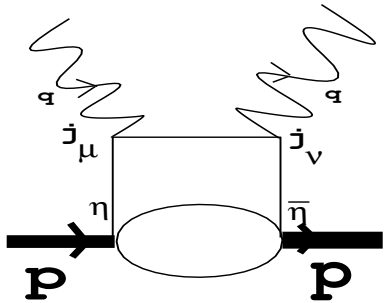 }}
  \caption{The lowest--order contribution to the four--point correlator
of Equation \protect\ref{4--point correlator}. }
  \label {lowest--order 4--point correlator}
\end{figure}

The determination of the functional dependence of the nuclear 
structure functions on the Bjorken scaling variable $x$ via QCD sum
rules has been pioneered by Belyaev and Ioffe
\cite{BelIof-dis1,BelIof-dis2}. They considered the four--point correlator
\begin{eqnarray}
\label{4--point correlator}
i T_{\mu\nu}(p,q) &=& \int\,d^4x\,d^4y\,d^4z\: e^{iqx} e^{ip(y-z)}
\langle T\{\eta(y) j_{\mu}(x) j_{\nu}(0)
\bar{\eta}(z)\}\rangle_0
\end{eqnarray} 
where $\eta$ is the standard three--quark current with proton quantum
numbers and $j_{\mu}$ is the electromagnetic current (see
Fig.~\ref{lowest--order 4--point correlator}). The sum rules they obtained have a limited applicability region
restricted to intermediate $x$ far from the kinematical boundaries
$x=0$ and $1$, and should be used for not very large $Q^2
\preceq 10\:GeV^2$.

\begin{figure}
  \epsfxsize=12.cm
  \epsfysize=5.cm
  \centerline{\epsfbox{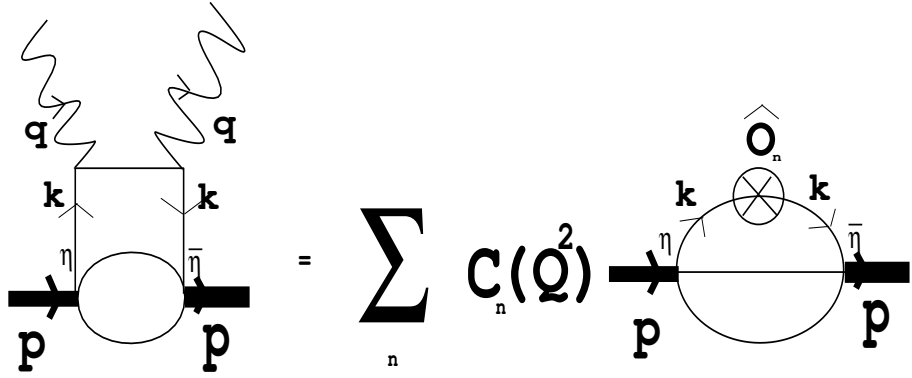 }}
  \caption{Applying the OPE for the upper propagator in the
four--point correlator approach leads to the three--point correlator approach.}
  \label {4-pt-to-3-pt}
\end{figure}

In this paper we propose another method to calculate the quark
distributions by evaluating their moments. More precisely, we shall
use QCD sum rules to determine the leading twist--two non--singlet
operator matrix elements (OMEs) contributing to deep inelastic
scattering from nucleon targets which, by the moments sum rules, are related to the quark distribution moments. In principle, this
method should be equivalent to that of \cite{BelIof-dis1} since it is based on
calculation of three--point correlation functions which can be
produced by performing the OPE on the two electromagnetic currents in the
four--point correlator approach (see Fig.~\ref{4-pt-to-3-pt}). However, it does not have the $Q^2$
problem as it is
calculated perturbatively from the perturbative dependence of the
coefficient functions and the anomalous dimension of the operators in
the OPE series.

It is crucial to note that, assuming the validity of the conventional
OPE as was the case in \cite{BelIof-dis1}, the calculations for higher
moments do not reveal logarithmic divergences. Thus it is not
necessary to use the external field method contrary to the case of the
second moments where the logarithmic divergences are present \cite{BelBlo2}. \footnote{In fact, performing the calculations in the case of the
second moment would lead to integrals of the form
$\int\,d^4x\,e^{ipx}\frac{\ln-x^2}{(x^2)^m}$ where $m\,>\,2$. These
integrals when Borel transformed give a divergent contribution.}
 
\section{The method}
Following the standard procedure of QCD sum rules, we calculate
the three--point correlation function
\begin{eqnarray}
T_{ij}(p)&=&\int d^{4}(x) e^{ipx} \langle
0|T\{\eta_{i}(x)\hat{O}_{\beta\alpha_{1}\ldots\alpha_{n}}\bar{\eta_{j}}(0)\}
|0 \rangle
\end{eqnarray} 
where $\eta_{i}(x)$ is the interpolating current for a proton
originally suggested in \cite{Ioffe-mass}.
\begin{eqnarray}
\eta_{i}(x)=(u^{T}_{a} (x)C\gamma_{\mu}u_{b}(x)) (\gamma_{5}
\gamma^{\mu} d_{c}(x))_{i} \epsilon^{abc}
\end{eqnarray}
$C$ is the charge conjugation matrix, superscript $T$ means transpose,
$i$ and $j$ are spinor indices while
$a$, $b$ and $c$ are colour indices. As to the operator $\hat{O}_{\beta\alpha_{1}\ldots\alpha_{n}}$, we
choose the twist--two non--singlet composite operator given by
\begin{eqnarray}
\hat{O}_{\beta\alpha_{1}{\ldots}\alpha_{n}}&=&i^{n} {\cal{S}} \int
d^{4}y\overline{\Psi}(y)\gamma_{\beta}D_{\alpha_{1}}{\ldots}D_{\alpha_{n}}\Psi(y)
\end {eqnarray}
where $D_\alpha$ is the covariant derivative
\begin{eqnarray}
\label{covariant derivative}
D_{\alpha}|_y \; =
\;\partial_{\alpha}-ig(\frac{\lambda^n}{2})A^n_{\alpha}(y) &=&
\partial_{\alpha}-i\frac{g}{2}(\frac{\lambda^n}{2})
y_{\rho}G^n_{\rho\alpha}(0)
\end {eqnarray}
in Schwinger gauge where $y_\alpha A^n_{\alpha}(y)=0$, and $\cal{S}$
denotes the symmetrization over the indices
$\beta,\alpha_{1},\ldots,\alpha_{n}$ with subtraction of traces in
order to make the operator $\hat{O}$ of definite spin.

The operators $\hat{O}_{\beta\alpha_{1}{\ldots}\alpha_{n}}$ form an
irreducible representation of the Lorentz group. Their reduced matrix
elements $A^n$ are defined by
\begin{eqnarray} 
\langle\,N(p)|\hat{O}_{\beta\alpha_{1}{\ldots}\alpha_{n}}
|N(p)\,\rangle &=& A_n\;p^{\beta}p^{\alpha_1}{\ldots}p^{\alpha_n}\;+\;
terms\; containing \; g_{\mu_i\mu_j}\,,
\label{OME: A_n}
\end {eqnarray}
and our aim is to estimate these OMEs' $A^n$ by the QCD sum rules
method.

First, we consider the ``phenomenological'' representation of the
three--point correlator which is expressed in terms of physical intermediate states as 
\begin{eqnarray}
\label{rhs 1}
&&{\langle}0|\eta_{i}|N(p)\rangle\times\;nucleon\;propagator\times
\langle\,N(p)|\hat{O}_{\beta\alpha_{1}{\ldots}\alpha_{n}}|N(p)\rangle\nonumber
\\
&&\times\;nucleon\;propagator\times\langle\,N(p)|\bar{\eta}_{j}|0\rangle
\end {eqnarray}
where $nucleon\;propagator\; = \; i \frac{\hat{p}+m}{p^2-m^2}$ and $m$ is the
mass of the nucleon $N$. In addition to this double pole term, there are
also single pole (continuum) terms suppressed
relative to the former with power of the Borel parameter
$M^{-2}$ (exponentially). One can check that the
combination of invariant functions which
enters the coefficient at the structure
$\hat{p}p^{\beta}p^{\alpha_1}{\ldots}p^{\alpha_n}$ or
$p^{\beta}p^{\alpha_1}{\ldots}p^{\alpha_n}$ in Equation~\ref{rhs 1} coincides (up to a
numerical factor) with the combination of invariant functions at the
structure $p^{\beta}p^{\alpha_1}{\ldots}p^{\alpha_n}$ in the spin--averaged matrix element $\langle\,N(p)|\hat{O}_{\beta\alpha_{1}{\ldots}\alpha_{n}}
|N(p)\,\rangle$ i.e. with the reduced matrix element we are interested
in, and we get:    
\begin{eqnarray}
\label{RHS}
Phen&=&\frac{-\lambda^2}{(p^2-m^2)^2}2A_n(\hat{p}+m)
p^{\beta}p^{\alpha_1}{\ldots}p^{\alpha_n}\;+\;other\: structures,
\end{eqnarray}
where $\lambda$ is the constant defined by
\begin{eqnarray}
{\langle}0|\eta|N(p)\rangle&=&\lambda v_p
\end{eqnarray}
and $v_p$ is the proton spinor satisfying $(\hat{p}-m)v_p=0$ and the
normalisation $\sum_{polarisations\;r}\bar{v}^rv^r=2m$.

The different tensor structures emerging from the double pole term can
be used to construct a sum rule and extract $A_n$, but in our case the sum rule from the structure
$\hat{p}p^{\beta}p^{\alpha_1}{\ldots}p^{\alpha_n}$ is
preferred. Firstly, this is because it contains the maximum number of momenta
in the numerator and thus improves the convergence of the OPE series and
diminishes the background contribution of excited hadronic states
compared to the lowest state (proton) contribution of
interest \cite{ioffe-proc}. Secondly, the structure
$\hat{p}p^{\beta}p^{\alpha_1}{\ldots}p^{\alpha_n}$ conserves
chirality: this is also a merit since for structures conserving
chirality one can calculate a larger number of terms in the
OPE series. Finally, for this tensor structure, several simplification tricks
proved to be possible which made the calculations for the
``theoretical'' part
manageable.

\begin{figure}
  \epsfxsize=12.cm
  \epsfysize=5.cm
  \centerline{\epsfbox{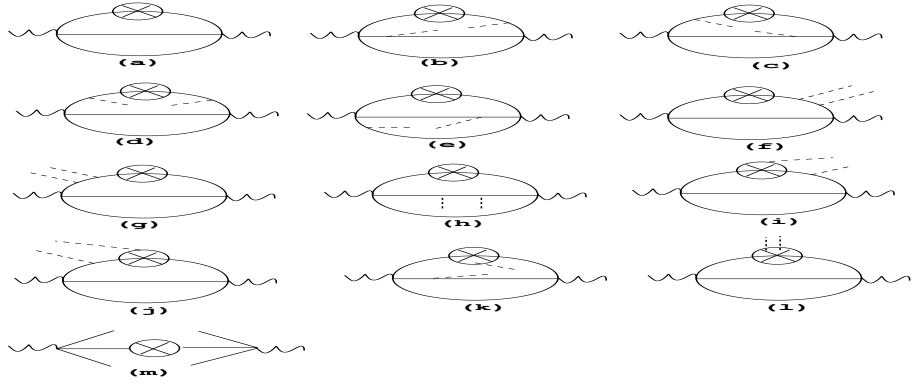 }}
  \caption{The ``theoretical'' sides of the sum rule for the
three--point correlation function. The wavy lines denote the current,
and the crossed circles denote the operator insertion. The
solid and dashed lines denote quarks and gluons respectively.  }
  \label {LHS diagrams}
\end{figure}

We consider now the ``theoretical'' part of the sum rules. In
the process of the calculations we shall take into account only the
operators of dimension $d\leq6$ and calculate only the invariant functions at the
structure $\hat{p}p^{\beta}p^{\alpha_1}{\ldots}p^{\alpha_n}$.
The diagrams corresponding to this tensor structure are depicted in
Fig.~\ref{LHS diagrams}
where the unit operator contribution corresponds to Fig.~\ref{LHS
diagrams}a, the gluon condensate ${\langle}
G^a_{\mu\nu}G^a_{\mu\nu}\rangle$ to Fig.~\ref{LHS diagrams}b--l and
the quark condensate $\langle\bar{\psi}{\psi}\rangle^2$ to
Fig.~\ref{LHS diagrams}m. For massless quarks, it is convenient to work in configuration
space and in order to calculate the gluonic condensate contribution it is far
easier to work in the fixed--point gauge for the gluonic field
$z_{\mu}A^n_{\mu}(z)=0$. 

\begin{figure}
  \epsfxsize=12.cm
  \epsfysize=4.cm
  \centerline{\epsfbox{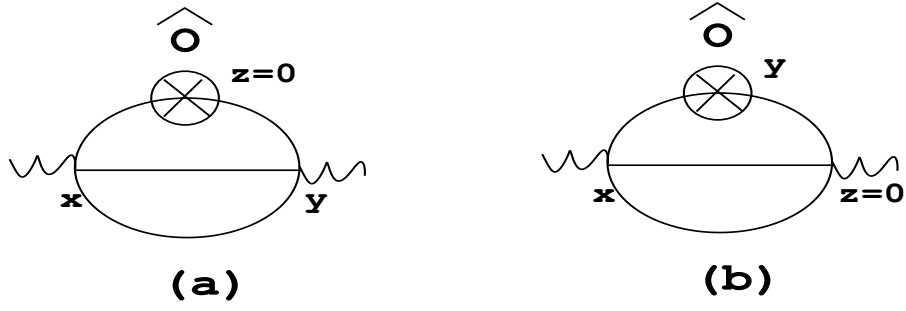 }}
  \caption{Two equivalent choices for the origin.}
  \label {two gauge choices}
\end{figure}

\begin{figure}
  \epsfxsize=12.cm
  \epsfysize=4.cm
  \centerline{\epsfbox{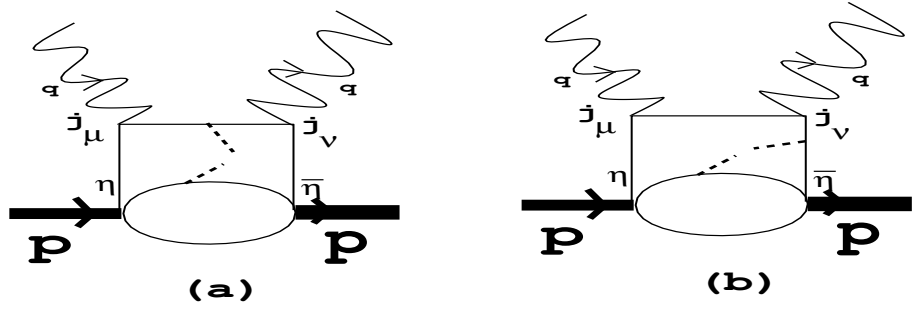 }}
  \caption{The contribution of graph (a) is of
higher twist while graph (b) may contribute to leading
twist effects.}
  \label {4-point-formalism}
\end{figure}

The standard choice of the reference frame is to identify the origin
($z=0$) with the coordinate of one of the currents, so we have the
following two
choices.
\begin{enumerate}
\item \underline{The operator $\hat{O}$ is inserted at the origin (Fig.~\ref {two gauge choices}a):}
\label{gauge choice a}
The advantage of this gauge is that, because covariant derivatives
may be replaced by ordinary derivatives at the origin, the graphs where gluons originate from the operator
(Fig.~\ref{LHS diagrams}i, j, k, l) do not contribute. This
simplification is, however, illusory and one should be very careful in
doing the calculations in this gauge choice. In fact, because the
operator contains derivatives, it is impossible to put ($z=0$) from
the very beginning since the quark loop is determined  now by
derivatives of the type $(\frac{\partial}{{\partial}z_\alpha})S(z,y)$
where $S(z,y)$ denotes the propagator between the two points $y$ and
$z$ and is given by \cite{review,NovShi,IofSmi-static}:
\begin{eqnarray}
S_{ij}^{A,B}(z,y)&=&
\delta ^{AB} \frac{i}{2\pi^2}\frac{(\widehat{z-y})_{ij}}{(z-y)^4}
+ 
\frac{-1}{16\pi^2}\frac{ig}{2}G^n_{\alpha\beta}
\left(\frac{\lambda^n}{2}\right)^{AB}
\frac
{\left(\widehat{z-y}\sigma_{\alpha\beta}+\sigma_{\alpha\beta}\widehat{z-y}\right)_{ij}}{(z-y)^2}
\nonumber\\&&
+ 
\frac{i}{2\pi^2}\frac{ig}{2}G^n_{\alpha\beta}
\left(\frac{\lambda^n}{2}\right)^{AB}
y_{\alpha} z_{\beta}\frac{\left(\widehat{z-y}\right)_{ij}}{(z-y)^4}
\nonumber\\&&
+
\frac{i}{48\pi^2}\left(\frac{ig}{2}\right)^2 
G^n_{\alpha\beta}G^m_{\alpha\beta}
\left(\frac{\lambda^n}{2}\frac{\lambda^m}{2}\right)^{AB}
(z^2y^2-(z.y)^2)\frac{(\widehat{z-y})_{ij}}{(z-y)^4}
+
\frac{-1}{12} \langle\bar{q}q\rangle \delta^{AB} \delta_{ij},
\nonumber\\&&
\end{eqnarray}
where $A,B$ are colour indices, $i,j$ are spinor indices, 
$\sigma_{\alpha\beta}=\frac{i}{2}\left[\gamma^{\alpha},\gamma^{\beta}\right]$
and 
$\lambda^n$ are the {\it Gell--Mann} matrices. 

Since one has to differentiate first with respect to $z$ and only
put $z=0$ afterwards, we can see that the ``non--translation--invariant'' terms
in $S(z,y)$ would give non-zero results leading to
new integrals to be done. 
\item \underline{The current $\bar\eta$ is inserted at the origin (Fig.~\ref{two gauge choices}b):}
\label{gauge choice b}
Here the derivatives are applied at $y$ and no derivatives in the
current $\bar\eta$ at $0$ so we can put $z=0$ from the
beginning. Looking again at the expression for $S(z,y)$ we see that the graphs in
Fig.~\ref{LHS diagrams}f and h contribute nothing but those of  
Fig.~\ref{LHS diagrams}i, j, k and l do contribute since the covariant derivative 
$D$ contains operators of the gluon field
(c.f. Equation~\ref{covariant derivative}).

At first glance, this gives rise to some doubts as one might suspect
that the gluons originating from the operator (e.g.~Fig.~\ref{LHS
diagrams}k) correspond to gluons emitted from the upper propagator in
the four--point correlation function formalism
(Fig.~\ref{4-point-formalism}a). If this were true then it would lead
to inconsistencies, because in Schwinger gauge the gluon field $A_\mu$
is expressed in terms of the field strength tensor $G_{\mu\nu}$ and the standard analysis of OPE in
DIS shows that this would be of higher twist effect while our operator is of leading twist!
However, as the choice of gauge in~\ref{gauge choice a} shows, such
higher twist operators do not contribute. Thus gauge invariance
requires that these operator--originated
gluons have their origin in the right--hand upper propagator
(Fig.~\ref{4-point-formalism}b) and no inconsistencies persist.
\end{enumerate}

We opt for the second choice of gauge: the origin at the right--hand
vertex of the diagrams of Fig.~\ref{LHS diagrams}, $y$ as the point
where the operator
$\hat{O}_{\beta\alpha_{1}{\ldots}\alpha_{n}}$ is inserted while $x$
is the point where the nucleon is created. For this gauge choice, the
calculations are easier and can be done in two steps. We do
the $y$--integration first to obtain expressions of ``full''
propagators in the presence of the operator
$\hat{O}_{\beta\alpha_{1}{\ldots}\alpha_{n}}$, leaving the
$x$--integration till the end. 

As an example, doing the $y$--integration in Fig.~\ref{two gauge
choices}b results in a full propagator

\begin{tabular}{p{4.cm}p{0.1cm}p{12.cm}}
\epsfxsize=4.cm
\epsfysize=2.cm
\epsfbox{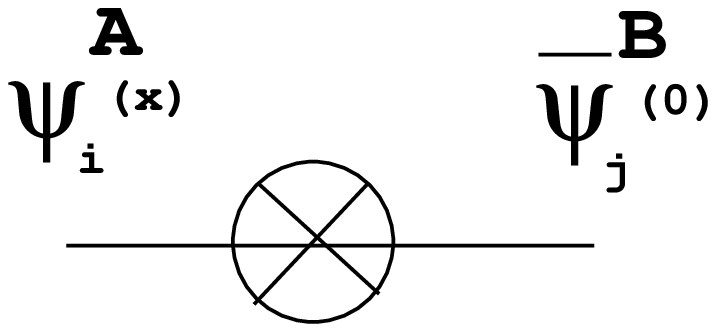} 
&=& 
$i^n\delta^{AB}{\cal{S}}{\int}d^4y\,\frac{i}{2\pi^2}\frac{\widehat{x-y}}{(x-y)^4}\;\gamma_{\beta}\partial_{\alpha_{1}}{\ldots}\partial_{\alpha_{n}}\;\frac{i}{2\pi^2}\frac{\hat{y}}{y^4}$
\\
\vspace{0.2cm}
&=&$i^n\delta^{AB}\frac{i}{2\pi^2}{\cal{S}}\partial_{\alpha_{1}}{\ldots}\partial_{\alpha_{n}}\left(\frac{\hat{x}x^{\beta}}{x^4}\right)$
\end{tabular}
where we have used the relation \cite{NovShi}
\begin{eqnarray}
\label{essential relation int p}
{\int}\frac{d^4p}{(p^2+i\epsilon)^n}e^{ip.x}&=&(-1)\frac{i(-1)^n2^{4-2n}\pi^2}{\Gamma(n-1)\Gamma(n)}(x^2-i\epsilon)^{(n-2)}\ln(-x^2+i\epsilon).
\end{eqnarray}

As we are interested only in the structure
$\hat{x}x^{\beta}x^{\alpha_1}{\ldots}x^{\alpha_n}$, the
derivatives can be done giving
  
\begin{tabular}{p{2.cm}p{0.05cm}p{12.cm}}
\epsfxsize=1.8cm
\epsfbox{free-propagator-inserted.eps}
&=&$i^n\delta^{AB}\frac{i}{2\pi^{2}}\left\{\frac{(-2)^{n-1}\:n!x^{\beta}}{(x^{2})^{n+1}}
\left(x^{\alpha_2}{\ldots}x^{\alpha_n}\gamma^{\alpha_1}+\ldots+x^{\alpha_1}{\ldots}x^{\alpha_{n-1}}\gamma^{\alpha_n}
\right)\right.$ 
\\&&$\left.\;\;\;\;\;\;\;\;\;\;\;\;\;\;\;\;
+\frac{(-2)^n(n+1)!}{(x^{2})^{n+2}}x^{\alpha_1}{\ldots}x^{\alpha_n}x^{\beta}\hat{x}\right\}$
\end{tabular}

Essentially the same techniques are used to evaluate other ``full''
propagators in the presence of operators even though the calculations
are more complicated. After this, one can evaluate the graphs: for
example, we get for the bare graph where the operator is inserted on a
u--quark line the result:

\begin{tabular}{p{4.cm}p{.1cm}p{12.cm}}
\vspace{0.05cm}
\epsfxsize=3.cm
\epsfbox{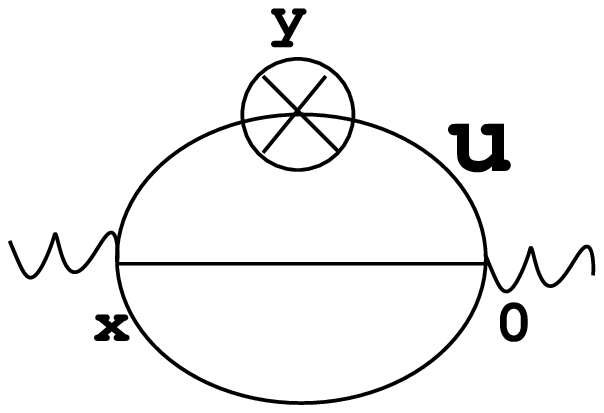} 
&\[=\]& \[\frac{i^{n+1}}{\pi^6}\frac{(-2)^nn!}{(x^2)^{n+5}}(-48-36n)\;\;.\]
\end{tabular}

In order to get the expression in the momentum representation, one
should now perform the $x$--integration using the formula 
\begin{eqnarray}
\label{essential relation int x}
{\int}\frac{d^4x}{(x^2-i\epsilon)^n}e^{ip.x}&=&\frac{i(-1)^n2^{4-2n}\pi^2}{\Gamma(n-1)\Gamma(n)}(p^2+i\epsilon)^{(n-2)}\ln(-p^2-i\epsilon)
\end{eqnarray}

For the other graphs, the calculations were performed using the program Mathematica for symbolic calculus
(we used the FeynCalc package). The results of all the graphs are
listed below (the letters a,b,$\ldots$ correspond to those of
Fig.~\ref{LHS diagrams}, the superscripts $u$ and $d$ denote that the operator is acting
on u-- and d--quarks respectively while the subscripts $u$ and $d$
mean that a u-- or d--quark line where the operator is not inserted
emits a gluon).
\begin{eqnarray}
(a)^u&=&\frac{1}{4\pi^4}\left(\frac{1}{2(n+1)}+\frac{3}{n+2}+\frac{-15}{2(n+3)}+\frac{4}{n+4}\right)p^2\ln-p^2
\\
(b\&c)^u_u&=&\frac{g^2{\langle}G^2\rangle}{\pi^4}\left(\frac{-1}{192(n+1)}+\frac{1}{192(n+2)}\right)\frac{1}{p^2}
\\
(b\&c)^u_d&=&\frac{g^2{\langle}G^2\rangle}{\pi^4}\left(\frac{1}{96(n+1)}+\frac{-1}{96(n+2)}\right)\frac{1}{p^2}
\\
(d)^u&=&\frac{g^2{\langle}G^2\rangle}{\pi^4}\left(\frac{-1}{432(n-1)}+\frac{1}{144n}+\frac{-1}{144(n+1)}+\frac{1}{432(n+2)}\right)\frac{1}{p^2}
\label{5_u}
\\
(e)^u&=&\frac{g^2{\langle}G^2\rangle}{\pi^4}\left(\frac{-1}{192(n+1)}+\frac{-1}{48(n+2)}\right)\frac{1}{p^2}
\\
(g)^u&=&\frac{g^2{\langle}G^2\rangle}{\pi^4}\left(\frac{-1}{144n}+\frac{-1}{192(n+1)}+\frac{5}{288(n+2)}\right)\frac{1}{p^2}
\\
(i\&j)^u&=&\frac{g^2{\langle}G^2\rangle}{\pi^4}\left(\frac{-1}{432(n-1)}+\frac{1}{192(n+1)}+\frac{-7}{864(n+2)}\right)\frac{1}{p^2}
\\
(k)^u_u&=&0
\\
(k)^u_d&=&\frac{g^2{\langle}G^2\rangle}{\pi^4}\left(\frac{-1}{192(n+1)}+\frac{1}{192(n+2)}\right)\frac{1}{p^2}
\\
(l)^u&=&\frac{g^2{\langle}G^2\rangle}{\pi^4}\left(\frac{-5}{576(n+1)}+\frac{1}{72(n+2)}\right)\frac{1}{p^2}
\\
(m)^u&=&0
\\
(a)^d&=&\frac{1}{4\pi^4}\left(\frac{1}{2(n+1)}+\frac{-3}{2(n+3)}+\frac{1}{n+4}\right)p^2\ln-p^2
\\
(b\&c)^d&=&\frac{g^2{\langle}G^2\rangle}{\pi^4}\left(\frac{-1}{64(n+1)}+\frac{1}{64(n+2)}\right)\frac{1}{p^2}
\\
(d)^d&=&\frac{g^2{\langle}G^2\rangle}{\pi^4}\left(\frac{-1}{432(n-1)}+\frac{1}{192n}+\frac{-1}{288(n+1)}+\frac{1}{1728(n+2)}\right)\frac{1}{p^2}
\\
(e)^d&=&\frac{g^2{\langle}G^2\rangle}{\pi^4}\left(\frac{-1}{192(n+1)}+\frac{1}{96(n+2)}\right)\frac{1}{p^2}
\\
(g)^d&=&\frac{g^2{\langle}G^2\rangle}{\pi^4}\left(\frac{-1}{192n}+\frac{1}{192
(n+1)}+\frac{1}{576(n+2)}\right)\frac{1}{p^2}
\\
(i\&j)^d&=&\frac{g^2{\langle}G^2\rangle}{\pi^4}\left(\frac{-1}{432(n-1)}+\frac{1}{192(n+1)}+\frac{-1}{216(n+2)}\right)\frac{1}{p^2}
\\
(k)^d&=&\frac{g^2{\langle}G^2\rangle}{\pi^4}\left(\frac{-1}{192(n+1)}+\frac{1}{192(n+2)}\right)\frac{1}{p^2}
\\
(l)^d&=&\frac{g^2{\langle}G^2\rangle}{\pi^4}\left(\frac{-1}{576(n+1)}+\frac{1}{288(n+2)}\right)\frac{1}{p^2}
\\
(m)^d&=&-\frac{4}{3}{\langle}\bar{q}q\rangle^2\frac{1}{p^4}
\end{eqnarray}

Summing all the graphs we arrive at: 
\begin{eqnarray}
\label{LHSu}
Theo^u&=&
\frac{4}{32\pi^4}\left(\frac{1}{n+1}+\frac{6}{n+2}+\frac{-15}{n+3}+\frac{8}{n+4}\right)\;p^2{\ln}-p^2\nonumber\\
&&+\;
\frac{g^2{\langle}G^2\rangle}{32\pi^4}\left(\frac{-4}{27(n-1)}+\frac{-2}{3(n+1)}+\frac{4}{27(n+2)}\right)\frac{1}{p^2}\\
Theo^d&=&
\label{LHSd}
\frac{4}{32\pi^4}\left(\frac{1}{n+1}+\frac{-3}{n+3}+\frac{2}{n+4}\right)\;p^2{\ln}-p^2\nonumber\\
&&+\;
\frac{g^2{\langle}G^2\rangle}{32\pi^4}\left(\frac{-4}{27(n-1)}+\frac{-2}{3(n+1)}+\frac{28}{27(n+2)}\right)\frac{1}{p^2}\nonumber\\
&&+\;\frac{-4}{3}{\langle}\bar{q}q\rangle^2\frac{1}{p^4}
\end{eqnarray}

Equating with Equation~\ref{RHS} and doing the Borel transform, we
arrive at
\begin{eqnarray}
\label{my-expression-u}
A_n^u + \frac{M^2}{m^2}C^u_n &=&
\frac{M^{6}}{2 \bar{\lambda}^{2}_{N}}e^{m^{2}/M^{2}}\left\{ 
4\left(\frac{1}{n+1}+\frac{6}{n+2}-\frac{15}{n+3}+\frac{8}{n+4}\right)
\right.\nonumber\\
&&+\left.\frac{b}{M^{4}} \left(
-\frac{4}{27}\frac{1}{n-1} -\frac{2}{3}\frac{1}{n+1} + \frac{4}{27}\frac{1}{n+2}
\right)\right\}\\
\label{my-expression-d}
A_n^d + \frac{M^2}{m^2}C^d_n &=&
\frac{M^{6}}{2 \bar{\lambda}^{2}_{N}}e^{m^{2}/M^{2}}\left\{ 
4\left(\frac{1}{n+1}+\frac{-3}{n+3}+\frac{2}{n+4}\right)
\right.\nonumber\\
&&+\frac{b}{M^{4}} \left(
-\frac{4}{27}\frac{1}{n-1} -\frac{2}{3}\frac{1}{n+1} + \frac{28}{27}\frac{1}{n+2}
\right)\nonumber\\
&&+\left.\frac{8}{3}\frac{a^2}{M^6}\right\}
\end{eqnarray}
where $m$ is the nucleon mass,
\[ a = -(2\pi)^{2}\langle 0 | \bar{\psi} \psi | 0 \rangle\;\;, \]
\[ b = (2\pi)^{2} \langle 0 | 	\frac{{\alpha}_ {s}}{\pi} {G^a}_{\mu\nu}{G^a}_{\mu \nu} 
	| 0 \rangle  \]
and \[ \bar{\lambda}^{2}_{N}\:=\: 2(2\pi)^4\lambda^2 \;\;.\]

\section{Comparison with other results}
As mentioned in the introduction, Belyaev and Ioffe \cite{BelIof-dis1} determined
the valence up and down quark distributions in the proton by
considering the four--point correlator $T_{\mu\nu}$ (Equation~\ref{4--point
correlator}) where $x=\frac{Q^2}{2p.q}$ is fixed and not too close to the boundaries
$x=0,1$ and $Q^2=-q^2$ is assumed to be only moderately large ($Q^2
\preceq 10\:GeV^2$) since the logarithmic
corrections $[\alpha_S(\ln\frac{Q^2}{\Lambda_{QCD}^2})]^n$ were not
summed. Assuming the legitimacy of using the conventional OPE for the
imaginary part of the correlator, they calculated
$Im\;T_{\mu\nu}$ taking perturbative and non--perturbative
corrections into account, performed the
Borel transformation over the parameter
$p^2$ keeping only leading powers in an expansion in
($\frac{1}{Q^2}$) and hence projecting only the leading twist
contribution. They obtained the following sum
rules (the effects of the continuum, the anomalous dimensions and the terms $\propto\, \alpha_{s}<\bar{q}q>^2$ are not
shown):
\begin{eqnarray}
\label{start1}
xu_{v}(x,Q^{2}) + M^{2} A^{u}(x,Q^{2}) &=&
\frac{M^{6}}{2 \bar{\lambda}^{2}_{N}}e^{m^{2}/M^{2}}  \left\{ 
4 x(1-x)^{2}(1+8x) \right.\nonumber\\
&&\left. +\frac{b}{M^{4}} \left(
-\frac{4}{27}\frac{1}{x}+\frac{7}{6}-\frac{19}{12}x+\frac{97}{108}x^{2}
\right) \right\}
\\
\label{start2}
xd_{v}(x,Q^{2}) + M^{2} A^{d}(x,Q^{2}) 
&=&\frac{M^{6}}{2 \bar{\lambda}^{2}_{N}}e^{m^{2}/M^{2}}
\left\{
4 x(1-x)^{2}(1+2x) \right. \nonumber \\
&&\left. +\frac{b}{M^{4}} \left(
-\frac{4}{27}\frac{1}{x}+\frac{7}{6}-\frac{11}{12}x-\frac{7}{54}x^{2}
\right) \right\}
\end{eqnarray}

The standard analysis of DIS would, in
principle, allow one to calculate the twist--two OMEs
simply by taking moments of the structure functions
\begin{eqnarray}
\label{defmomentu}
A_n^u	&=& \int_{0}^{1}dxx^nu_{v}(x,Q^2)\\
\label{defmomentd}
A_n^d	&=& \int_{0}^{1}dxx^nd_{v}(x,Q^2)
\end{eqnarray}
and formally one obtains
\begin{eqnarray}
\label{umoment}
A_n^u + \frac{M^2}{m^2} R^{u}_n 
&=&\frac{M^{6}}{2 \bar{\lambda}^{2}_{N}}e^{m^{2}/M^{2}}
 \left\{ 
4
\left(\frac{1}{n+1}+\frac{6}{n+2}-\frac{15}{n+3}+\frac{8}{n+4}\right)
\right.\nonumber\\
&&\left. +\frac{b}{M^{4}} \left(
-\frac{4}{27}\frac{1}{n-1} + \frac{7}{6}\frac{1}{n}
-\frac{19}{12}\frac{1}{n+1} + \frac{97}{108}\frac{1}{n+2}
\right)\right\}
\\
\label{dmoment}
A_n^d + \frac{M^2}{m^2} R^{d}_n
&=& \frac{M^{6}}{2 \bar{\lambda}^{2}_{N}}e^{m^{2}/M^{2}}
\left\{ 
4
\left(\frac{1}{n+1}+\frac{-3}{n+3}+\frac{2}{n+4}\right) \right. \nonumber\\
&&+\frac{b}{M^{4}} \left(
-\frac{4}{27}\frac{1}{n-1} + \frac{7}{6}\frac{1}{n}
-\frac{11}{12}\frac{1}{n+1} + \frac{-7}{54}\frac{1}{n+2}
\right)\nonumber \\
&&\left. + \frac{8}{3}\frac{a^2}{M^6}  \right\}.
\end{eqnarray}
The term proportional to $a^{2}$ in Equation~\ref{dmoment}
corresponds to the $\delta(1-x)$ piece which was omitted in the
structure function expression in Equation~\ref{start2} applicable at
intermediate values of $x$. 

\section{Discussion}
In comparing Equations~\ref{my-expression-u} and~\ref{my-expression-d}
with Equations~\ref{umoment} and~\ref{dmoment}, 
we see that the unit operator and $<\bar{q}q>^2$ contributions agree
while there is a discrepancy in the $<G^2>$ term. In Belyaev's opinion
\cite{Bel-emails}, it is not necessary that
(\ref{my-expression-u}, \ref{my-expression-d}) and
(\ref{umoment}, \ref{dmoment}) give the same answer because the sum
rules (\ref{start1}, \ref{start2}) are used to find the quark
distribution for intermediate $x$ only, and one cannot integrate them
analytically over the whole region $x\in[0,1]$ because there can be
singular contributions of the OPE near the points $x=0,1$.
However, in a formal (rather than phenomenological) sense, one can compare the two methods. Since we can identify the
singularities at $x=0,1$ in the approach of \cite{BelIof-dis1}, we can
formally evaluate the moments and compare them with the OPE analysis. We have not been
able to establish the origin of the discrepancy although, as was noted in \cite{thesis}, the phenomenological analysis
of the OMEs \cite{nidal1,paper1} changes only slightly using the new $<G^2>$ term
expression. However  the difference is still of importance and we
think it essential to understand the origin of the discrepancy in the two
methods. We hope this paper will stimulate further investigation of
this puzzle.

{\bf Acknowledgement:} I am very grateful to G.G. Ross for help in
this analysis.


\begin{thebibliography}{99}
\bibitem{BelIof-dis1}V.M. Belyaev and B.L. Ioffe, Nucl. Phys. B310
(1988) 548.
\bibitem{BelIof-dis2}V.M. Belyaev and B.L. Ioffe,
Int. J. Mod. Phys. A6 (1992) 1533.

\bibitem{BelBlo2}V.M. Belyaev and B.Yu. Blok, Z. Phys. C - Particles
and Fields 30 (1986) 279.
\bibitem{Ioffe-mass}B.L. Ioffe, Nucl. Phys. B188 (1981) 317; Erratum:
Nucl. Phys. B191 (1981) 591.
\bibitem{ioffe-proc}B.L. Ioffe, Proc. of XXII Int. Conf. on
High Energy Physics, Leipzig (1984), vol. 2,  176.
\bibitem{review}L.J. Reinders, H. Rubinstein and S. Yazaki,
Phys. Rep. 127 (1985) 1.
\bibitem{NovShi}V.A. Novikov, M.A. Shifman, A.I. Vainshtein and V.I. Zakharov,
Fortschr. Phys. 32 (1984) 600.
\bibitem{IofSmi-static}B.L. Ioffe and A.V. Smilga, Nucl. Phys. B232
(1984) 109.
\bibitem{Bel-emails}V.M. Belyaev, private communication.
\bibitem{thesis}N. Chamoun, Thesis (1996), OUTP--96--67P.
\bibitem{nidal1} G.G. Ross and N. Chamoun, Phys. Lett. B380 (1996) 151.
\bibitem{paper1} N. Chamoun and G.G. Ross, in preparation. 
\end{thebibliography}
\end{document}